# Injection locking of Rydberg dissipative time crystals

Darmindra Arumugam

Jet Propulsion Laboratory, California Institute of Technology, Pasadena, 91109, California, USA

E-mail: darmindra.d.arumugam@jpl.nasa.gov;

**Abstract**

Non-equilibrium Rydberg gases exhibit exotic many-body phases stabilized by the interplay of coherent interactions and dissipation. Strong Rydberg interactions drive sustained limit cycle oscillations, whose robustness, long-range temporal order, and spontaneous time-translation symmetry breaking establish a dissipative time crystal (DTC). Collective self-entrainment in driven ensembles leads to global synchronization and a dominant oscillation frequency. Here, injection locking of a Rydberg DTC is demonstrated using a radio-frequency (RF) electric field that gradually pulls the intrinsic oscillation toward the injected frequency. Above a critical threshold, full synchronization occurs, with the locking bandwidth scaling linearly with RF amplitude. This includes synchronization of higher-order harmonics, revealing entrainment of the system's nonlinear temporal dynamics. The phenomenon parallels injection locking in classical nonlinear systems, but emerges here in a strongly interacting quantum medium. This approach establishes a new method for stabilizing and controlling quantum temporal order, with applications in precision sensing, quantum metrology, and timekeeping.

Driven-dissipative Rydberg ensembles offer a rich platform for exploring emergent many-body phases[1-5]. At room temperature and under weak magnetic fields, mode competition between closely spaced Rydberg sublevels leads to self-sustained oscillations[6] (OSC)—hallmarks of a dissipative time crystal (DTC)[7-16] with broken time-translation symmetry and long-range temporal order. Here, a radio-frequency (RF) electric field (E-field) is used to perturb this emergent OSC phase. As the drive RF nears the intrinsic oscillation frequency, the system undergoes frequency pulling and synchronizes above a critical threshold—demonstrating injection locking[17,18] in a strongly interacting quantum Rydberg medium. Although injection locking via external fields has been demonstrated in quantum systems[19,20], it has not been realized in Rydberg many-body phases, where strong interactions and dissipation enable new synchronization dynamics. This highly controllable entrainment mechanism opens a path to stabilize the Rydberg DTCs temporal order with applications in sensing, metrology, and timekeeping.

The locking bandwidth, $\Delta\omega_{\text{lock}}$, is experimentally observed by sweeping the RF frequency and amplitudes to identify the frequency range for synchronization. It is found to increase linearly with the RF field, following $\Delta\omega_{\text{lock}}=\kappa E$, where $\kappa$ is a forcing strength and $E$ the RF field amplitude. The mean-field theory with RF-induced coherence between excited states captures the locking dynamics. Experimental observations demonstrate synchronization of higher-order harmonics, revealing full entrainment of the system's nonlinear temporal dynamics. Together, these results provide a quantitative framework for externally entraining temporal order in an interacting Rydberg atomic many-body system.

## Results

**Injection pulling and synchronization dynamics**

The Rydberg dissipative time crystal[6] (DTC) exhibits spontaneously generated oscillations (OSC) arising from internal mode competition[4-6] between closely spaced Zeeman-shifted Rydberg sublevels[6]. These limit cycle oscillations form a self-organized temporal order that breaks continuous time-translation symmetry (see Supplementary Section 1). When an external periodic RF drive is introduced, the system's intrinsic oscillations (OSC) undergo frequency pulling and can gradually synchronize to the RF drive — a phenomenon analogous to injection locking in classical nonlinear dynamical systems[17,18,21-24]. This behavior mirrors the predictions of Adler[17] (see Methods) where phase coupling between the natural oscillator and external forcing governs the transition from free-running to locked dynamics through a restoring torque in phase space.

To probe this synchronization behavior, a weak RF E-field is applied across a room-temperature cesium vapor cell (Cs-133) using a parallel-plate capacitor (Fig. 1a). The atoms are excited via a ladder-type scheme: a probe laser at 852 nm is detuned by $\Delta_p/2\pi$ = +22 MHz from the $6S_{1/2} \rightarrow 6P_{3/2}$ transition, while a coupling laser at 508 nm drives $6P_{3/2} \rightarrow 76D_{5/2}$ with $\Delta_c/2\pi$ = -30 MHz (Fig. 1b). A static magnetic field (B = 4 G) introduces Zeeman shifts among the Rydberg sublevels, enabling the emergence of spontaneous oscillations (OSC) in probe transmission[6]–which arise due to nonlinear mode competition (see Supplementary Section 1). Details of the experimental setup, protocols, and systems is given in Methods and Supplementary Section 2. The applied RF E-field introduces a weak coherent coupling between the excited Rydberg states |r⟩ and |s⟩ (Fig. 1b), detuned by $\delta_s$ from their energy separation. Operating near the system's natural OSC frequency, it modulates the phase of the excited-state coherence (see Methods).

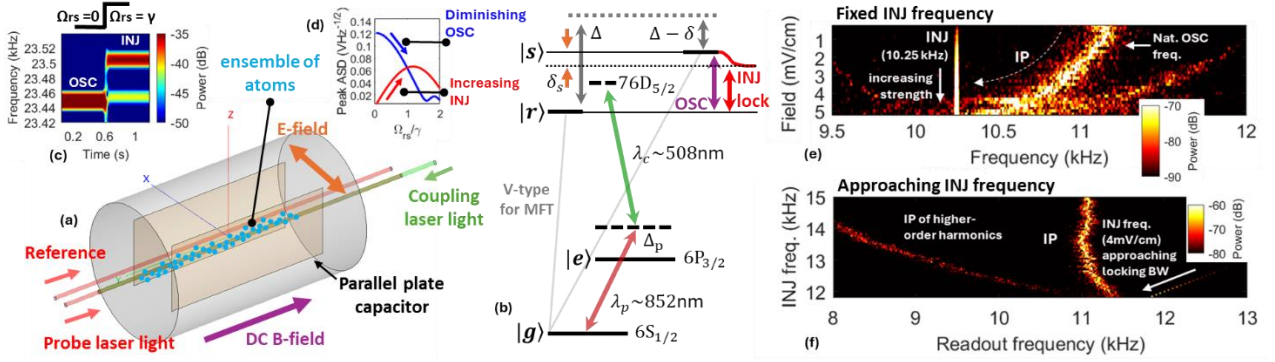

**Fig. 1: Experimental protocol, energy diagram, mean-field treatment simulations, and experimental observations. (a)** Counter-propagating probe and coupler laser beams are directed through a room-temperature Cesium vapor cell. Inside the cell, a parallel-plate capacitor enables the application of low-frequency electric fields (E-field). A static magnetic field (B-field) is applied to induce mode competition among the excited Rydberg state sublevels, resulting in limit cycle oscillations (OSC). To reduce classical noise, a reference probe beam is employed. **(b)** The experimental setup follows a ladder-type configuration, where a probe laser at λp∼852nm drives the ground state |g⟩=6S$_{1/2}$ to the intermediate excited state |e⟩=6P$_{3/2}$, detuned by Δp/2π=22MHz. A coupler laser at λc∼508nm then excites atoms from |e⟩ to a high-lying Rydberg state 76D$_{5/2}$, detuned by Δc/2π=30MHz. A B=4G field is applied to induce Zeeman shifts and drive competition between the Rydberg sublevels. A low-frequency E-field modulates the energy difference between two nearby Rydberg states |r⟩ and |s⟩, with the detuning δs defined relative to their energy separation. For theoretical modeling, a V-type three-level system (gray) is used to approximate the dynamics. **(c)** Time-dependent numerical simulations of the V-type model shows that the OSC frequency is dynamically frequency locked to injection (INJ) signal when turned on. **(d)** Numerical simulation with varying INJ Rabi frequency shows a peak INJ lock close to Ωrs/γ∼1, where γ is the linewidth of the upper states. **(e)** Experimental observations for fixed RF E-field frequency (f$_{INJ}$=10.25kHz) shows that as E-field magnitude is increased, the OSC (naturally at ∼11.2kHz) is gradually pulled towards the INJ signal until at abruptly locks to the INJ frequency. This is the characteristic injection pulling (IP) in oscillatory dynamics. **(f)** For a fixed E-field magnitude (E=4mV/cm), a similar gradual IP pulling is observed when the RF E-field frequency is gradually moved towards the natural OSC frequency.

Numerical simulations based on a simplified V-type mean-field treatment (Fig. 1c) demonstrate that the frequency of the natural OSC locks to the injected RF drive (INJ) once the Rabi frequency $\Omega_{rs}$ approaches the decay rate ($\Omega_{rs}/\gamma \sim 1$, Fig. 1d). Experimentally, injection pulling (IP) is observed when the INJ field is tuned near the intrinsic OSC frequency ($f_{OSC} \sim 11.2$ kHz). As shown in Fig. 1e, increasing the RF field strength at a fixed $f_{INJ}$ = 10.25 kHz causes the OSC to shift gradually toward the drive frequency until it abruptly locks—a clear signature of injection locking. Similarly, sweeping $f_{INJ}$ toward $f_{OSC}$ at fixed field strength (E = 4 mV/cm) produces a comparable pulling and locking behavior (Fig. 1f).

The emergence of INJ locking is further quantified in Fig. 2a–c. Transmission spectra are recorded while sweeping $f_{INJ}$ towards $f_{OSC}$ at different field strengths. At strong drive (E = 4.2 mV/cm), the OSC frequency smoothly shifts toward $f_{INJ}$ and locks near 11.8 kHz (Fig. 2a). At weak drive (E = 0.83 mV/cm), no frequency shift occurs, indicating that synchronization does not occur below a threshold (Fig. 2b). At E = 5.8 mV/cm, locking is observed at a higher injected frequency of ~12.3 kHz (Fig. 2c). Color-mapped spectra in Fig. 2d–f visualize the dependence of locking on both the injected frequency and field strength. The minimum RF amplitude required to achieve locking increases with detuning between $f_{INJ}$ and $f_{OSC}$, consistent with a locking bandwidth that scales linearly with the drive amplitude[17,18]. This behavior is in agreement with a phase model derived in the Methods section, where locking occurs when the detuning $\Delta \omega$ between $f_{INJ}$ and $f_{OSC}$ falls within a bandwidth $\Delta \omega_{\text{lock}} = 2K\Omega_{rs}^{\text{crit}}$, where $\Omega_{rs}^{\text{crit}}$ is the critical INJ Rabi frequency and K is the theoretical forcing strength scaling factor.

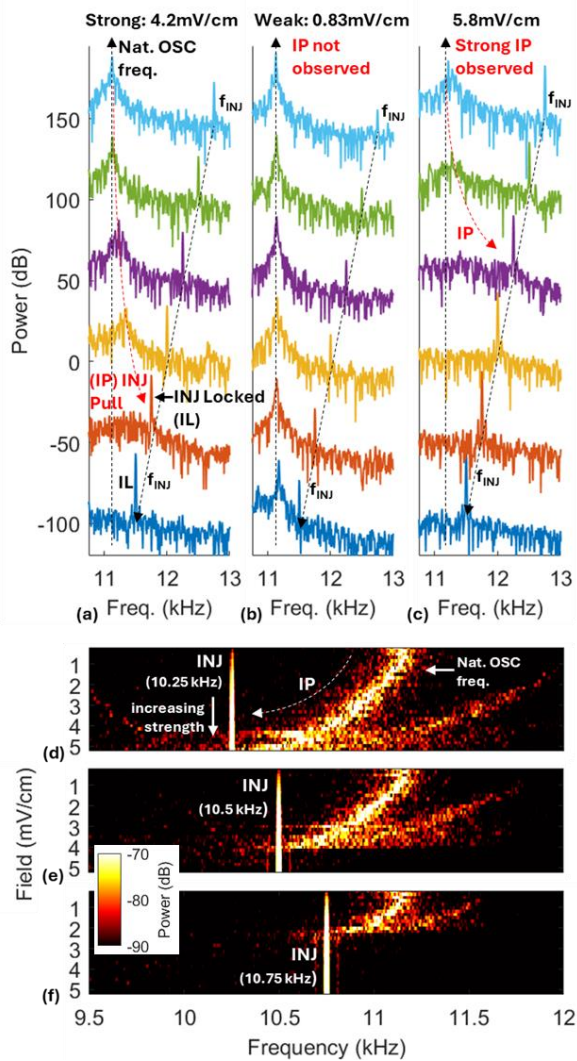

**Fig. 2:** (a-c) Spectral snapshots of OSC dynamics under strong and weak RF injection (INJ). The INJ frequency ($f_{INJ}$) is gradually moved towards the natural OSC frequency. (a) E=4.2mV/cm. OSC at about 11.2kHz is seen to gradually shift towards INJ, locking at ~11.8kHz. The locking is referred to as INJ lock (IL). Within an effective bandwidth from the OSC, referred to as the locking bandwidth, the signal remains locked. (b) E=0.83mV/cm. OSC at about 11.2kHz remains at the natural OSC frequency and does not shift towards INJ, demonstrating that weak INJ do not result in IP (pulling). (c) E=5.8mV/cm. OSC at about 11.2kHz rapidly shifts towards INJ, locking at ~12.3kHz. The observations confirm that IP is dependent on the strength of the INJ signal, and that the locking bandwidth is dependent on INJ field magnitude. (d-e) Colormap datasets of INJ field magnitude vs spectral frequency showing dependence of IL as a function critical INJ field magnitude. Natural OSC frequencies at about 11.2kHz in all cases. (d) $f_{INJ}$= 10.25kHz. IL occurs at ~5mV/cm. (e) $f_{INJ}$= 10.5kHz. IL occurs at ~4mV/cm. (f) $f_{INJ}$= 10.75kHz. IL occurs at ~2.5mV/cm. The critical field magnitude to IL the OSC increases as a function of separation between INJ and OSC frequencies.

Additional insight into the nonlinear response of the system is provided in Fig. 3. As $f_{INJ}$ is swept toward $f_{OSC}$ at fixed field amplitude (E = 4 mV/cm), not only the fundamental oscillation but also its higher harmonics undergo pulling toward the injected signal (Fig. 3a), indicating entrainment of the full nonlinear temporal dynamics. In a step-on protocol (Fig. 3b), when the RF drive is suddenly activated within the locking bandwidth, the system rapidly locks to the injected frequency, confirming deterministic frequency or phase capture. Frequency stabilization is further evidenced in Fig. 3c, where the injection-locked state can exhibit reduced spectral-width (-3 dB bandwidth) and offers significantly reduced frequency drift compared to the free-running Rydberg DTC natural OSC, demonstrating robust external control of temporal order.

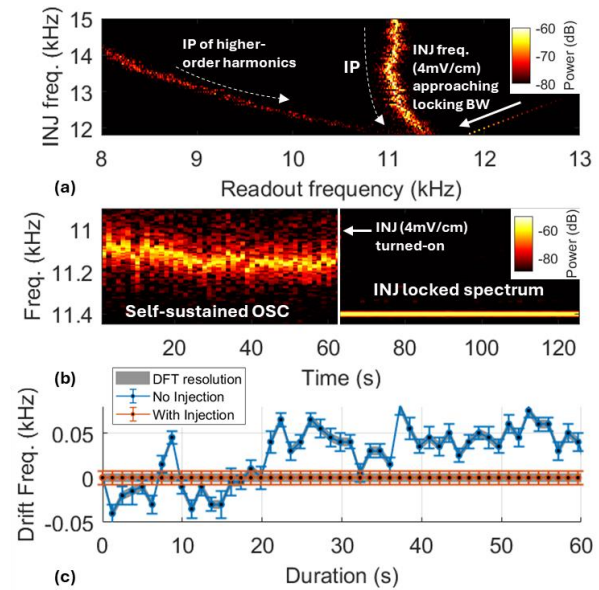

**Fig. 3:** (a) With E-field held constant at 4 mV/cm, a gradual pulling (IP) occurs as the RF field frequency is slowly tuned closer to the natural OSC frequency. Higher order harmonics of the OSC is also IP towards the INJ frequency. (b) In a step-on regime, where the INJ (4mV/cm) is turned on abruptly and is within the locking bandwidth, the OSC rapidly locks to the INJ. (c) Natural OSC dynamics show frequency drifts and have larger 3dB (FWHM) bandwidth (error bars) compared to after injection locking, where the dynamics and FWHM follow the characteristics of the INJ.

### Locking bandwidth and field scaling

To quantify synchronization strength, we extract the locking bandwidth—the frequency range over which the intrinsic oscillation remains phase-locked to the injected RF signal. As shown in Fig. 4a, sweeping the injection frequency from 12.3 to 10 kHz at a fixed E-field amplitude (4.2 mV/cm) reveals frequency pulling followed by locking near 11.8 kHz. Locking persists until the drive falls below ~10.6 kHz, yielding an estimated locking bandwidth of ~1.2 kHz.

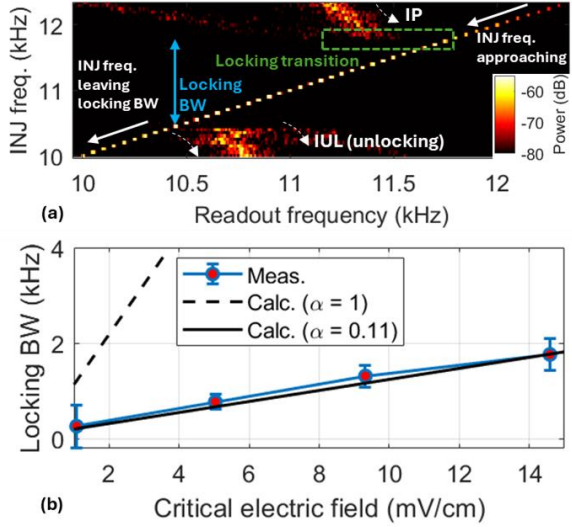

**Fig. 4: Locking, unlocking, and bandwidths. (a)** INJ frequency is swept from about 12.3-10kHz (4.2mV/cm). IP is observed to pull OSC towards INJ and locking occurs at about ~11.8kHz. The system remains IL (INJ locked) between ~11.8-10.6kHz (locking bandwidth is visually estimated here at about ~1.2kHz). **(b)** Locking bandwidth as a function of critical electric field, extracted by performing linear fits to the readout OSC frequency shift versus applied field for each injection frequency. The critical field is defined as the intercept point of the fit, corresponding to where the OSC is estimated to linearly intercept the INJ frequency. Experimental measurements (blue) are compared for two values of the empirical scaling factor α, where Δω= κE, κ=α(Kd$_{ind}$/ℏ), d$_{ind}$ is the induced dipole moment and K is the forcing strength scaling factor, and Δω is the locking bandwidth. The estimated Δω is reduced relative to visual estimation due to linear fit estimation of intercept points.

To extract the locking bandwidth, injection frequency is swept at fixed E-field amplitudes, and the oscillator's frequency pulling is tracked prior to synchronization. Linear fits to the readout frequency versus injection frequency on both sides of the locking region yield synchronization onset points. The difference between these intercepts defines the locking bandwidth, which is evaluated as a function of field strength.

As shown in Fig. 4b, the bandwidth increases linearly with the applied field, consistent with the injection-locking phase model prediction $\Delta\omega_{lock} = K\Omega_{rs}$. To connect this to experiment, we express the Rabi frequency as $\Omega_{rs} \sim d_{ind}E/\hbar$, yielding $\Delta\omega_{lock} = \kappa E$, where $\kappa = Kd_{ind}/\hbar$ defines the effective field-based scaling. The induced dipole moment $d_{ind}$ here can be through of as arising from weak magnetic-field mixing between two Rydberg levels — one of which is dipole-allowed. This mixing is treated perturbatively: the applied magnetic field admixes a small component of a dipole-connected state |r′⟩ into |r⟩, inducing a weak electric dipole moment. The effective coupling is estimated to scale as $d_{ind} \sim (H_B/\Delta E)d_{ref}$, where $H_B = \mu_B g_J m_J B$ is the Zeeman interaction energy, $\Delta E$ is the energy gap between |r⟩ and |r′⟩, and $d_{ref}$ is the dipole matrix element for the nearby allowed transition. The nearby allowed transition ($|76D_{5/2}\rangle \leftrightarrow |77P_{3/2}\rangle$) and B = 4G was used to estimate the induced dipole moment of $d_{ind} \sim 4.96 \times 10^{-28}$ C·m. To account for uncertainties in the mixing strength and interaction geometry, we introduce an empirical correction factor α, giving $\kappa \to K(\alpha d_{ind})/\hbar$. The observed linear trend in Fig. 4b is consistent with the model prediction, with $\alpha \sim 0.11$ indicating reduced effective coupling relative to the uncorrected mixing estimate.

### Discussion

The results presented here establish that dissipative time crystals (DTCs) based on strongly interacting Rydberg ensembles can be injection-locked using weak RF E-fields, providing a new mechanism for externally stabilizing this emergent temporal order. The transition from free-running oscillations to synchronized dynamics follows the canonical signatures of nonlinear phase entrainment, with behavior quantitatively captured by a mean-field V-type model (see Methods) and a phase evolution equation akin to Adler's equation[17]. Frequency pulling and abrupt locking are observed when the RF field strength exceeds a threshold, with the locking bandwidth scaling linearly with field amplitude, consistent with the prediction $\Delta\omega_{lock}=\kappa E$.

A key finding is the synchronization of not only the fundamental oscillation but also higher-order harmonics, indicating that the injected signal entrains the full nonlinear temporal structure of the DTC. This reflects the collective nature of the dynamics and highlights the ability of weak external fields to control complex quantum temporal phases. The experimentally extracted scaling constant $\kappa$ is described by a model incorporating magnetic-field-induced mixing between Rydberg states, with an empirical correction factor $\alpha \sim 0.11$. The agreement between measurement and theory supports the picture that even weak magnetic admixtures can enable effective coupling in otherwise dipole-forbidden transitions.

These findings demonstrate that DTC oscillations can be externally steered in frequency, narrowed in linewidth, and stabilized in phase via injection locking. The approach offers a tunable, low-noise route for synchronizing room-temperature quantum systems, relevant in E-field sensing, RF synchronization, and temporal metrology. In future work, this technique could be extended to arrays of interacting atomic DTC oscillators, exploration of collective synchronization phenomena, phase clustering, or topological time-domain ordering. Furthermore, by introducing engineered disorder or periodic modulation, it may be possible to realize more complex Floquet time crystal phases or investigate noise-induced synchronization thresholds in strongly dissipative media.

## Methods

### Injection dynamics in Rydberg mean-field models

The V-type atomic structure has been shown to be effective at developing a simplified framework to study natural OSC from Rydberg DTCs[6]. This motivates its use in advancing a framework to study injection locking from driven RF (injected) signals that are near the natural OSC from Rydberg DTCs. The analysis starts by constructing a representative Hamiltonian for a V-type system consisting of three energy levels (illustrated in Fig. 1b), where an external RF field introduces coupling between the excited states |r⟩ and |s⟩. This RF field is not perfectly resonant but instead operates with a slight detuning from the energy separation between those states. In the rotating frame, the Hamiltonian takes the form:

$$\hat{H} = \frac{\Omega}{2}\sum_i \left(\hat{\sigma}_{gr}^i + \hat{\sigma}_{gs}^i + \text{H.c.}\right) - \sum_i (\Delta_r \hat{n}_r^i + \Delta_s \hat{n}_s^i)$$
$$+ \frac{1}{2}\sum_{i\neq j} V_{ij}\left(\hat{n}_r^i \hat{n}_r^j + 2\hat{n}_r^i \hat{n}_s^j + \hat{n}_s^i \hat{n}_s^j\right)$$
$$+ \frac{\Omega_{rs}}{2}\sum_i \left(\hat{\sigma}_{rs}^i e^{i\delta_s t} + \hat{\sigma}_{sr}^i e^{-i\delta_s t}\right)$$

The operator $\hat{\sigma}_{\alpha\beta}^i = |\alpha^i\rangle\langle\beta^i|$ (with $\alpha,\beta = g,r,s$) is the transition between internal states at site $i$, while $\hat{n}_\alpha^i = |\alpha^i\rangle\langle\alpha^i|$ (for $\alpha = r,s$) corresponds to the local population in the Rydberg states. $\Omega$ denotes the Rabi frequency for transitions between ground and excited states, $\Omega_{rs}$ the coupling between excited states |r⟩ and |s⟩, and $\delta_s$ is the detuning frequency of the injected RF field. The interaction strengths $V_{ij}$ are taken to have uniform magnitude across all interacting pairs[4-6]. The behavior of the system is dictated by population dynamics that account for the interplay between excitation, spontaneous decay, and coherence effects (first-moment equations of motion for population and coherence are presented in Supplementary Section 3). Under the mean-field approximation, inter-atomic correlations are ignored, allowing high-order moment factorization — $\langle \hat{n}_r^i \hat{\sigma}_{gr}^j \rangle \approx \langle \hat{n}_r^i \rangle \langle \hat{\sigma}_{gr}^j \rangle$, $\langle \hat{n}_s^i \hat{\sigma}_{gs}^j \rangle \approx \langle \hat{n}_s^i \rangle \langle \hat{\sigma}_{gs}^j \rangle$. With uniform distribution $\langle \hat{n}_r^i \rangle = n_r$ and $\langle \hat{n}_s^i \rangle = n_s$, the interaction-induced nonlinear energy shift becomes $E_{NL} = \chi(n_r + n_s)$, where $\chi$ is the strength of atomic interactions. The mean-field population and coherence equations are given by (see Supplementary Section 3):

$$\dot{n}_r = \frac{\Omega}{2}\Im(\sigma_{gr}) - \gamma n_r + \frac{\Omega_{rs}}{2}\Im(\sigma_{rs}e^{i\delta_s t})$$
$$\dot{n}_s = \frac{\Omega}{2}\Im(\sigma_{gs}) - \gamma n_s - \frac{\Omega_{rs}}{2}\Im(\sigma_{rs}e^{i\delta_s t})$$
$$\dot{\sigma}_{gr} = i\frac{\Omega}{2}(2n_r + n_s + \sigma_{sr} - 1)$$
$$\qquad\quad + i\left(\Delta_r - E_{NL} + i\frac{\gamma}{2}\right)\sigma_{gr}$$
$$\dot{\sigma}_{gs} = i\frac{\Omega}{2}(2n_s + n_r + \sigma_{rs} - 1)$$
$$\qquad\quad + i\left(\Delta_s - E_{NL} + i\frac{\gamma}{2}\right)\sigma_{gs}$$

$$\dot{\sigma}_{rs} = i\frac{\Omega}{2}(\sigma_{gs} - \sigma_{rg}) - i(\Delta_r - \Delta_s - i\gamma)\sigma_{rs}$$
$$+ i\frac{\Omega_{rs}}{2}(n_r - n_s)e^{i\delta_s t}.$$

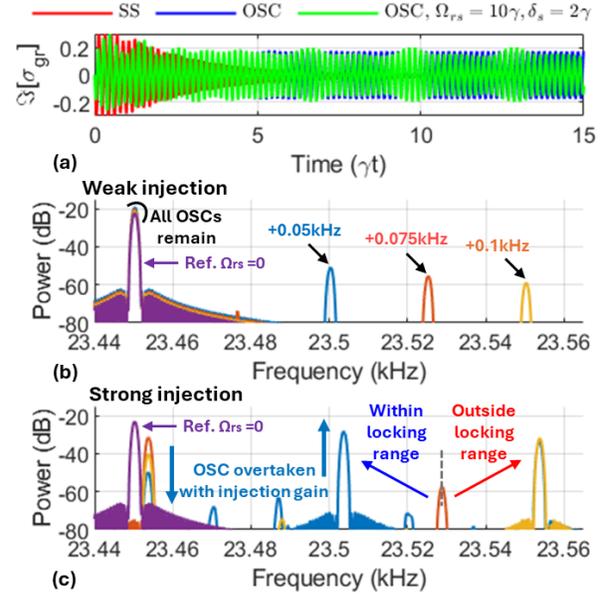

**Fig. 5:** Time-dependent numerical simulations of the V-type atom with a mean field treatment. (a) Transient response of the $\Im[\sigma_{gr}]$ that relates to probe transmission showing rapid decay (red) outside the OSC regime (stationary phase, SS), sustained oscillations (blue) in the OSC region, and perturbations when an RF is injected resulting in a gradual shift in the periodicity or frequency (green). (b-c) The spectrum of $\Im[\sigma_{gr}]$ is studied for weak and strong INJ signals with frequencies $f_{INJ}$ at +0.05kHz, +0.75kHz, and +0.1kHz relative to natural OSC frequencies. A reference line (purple) is included to identify the case where $\Omega_{rs}=0$ (no INJ signal). (b) Weak INJ signals do not perturb the OSC dynamics (at about ~23.45kHz) and increases in magnitude as it approaches the OSC (compare to measurements in Fig. 2b). (c) Strong INJ signals perturb the OSC. When INJ signals are within the locking bandwidth and above a critical magnitude, they result in a rapidly diminishing magnitude of the OSC signal at the natural OSC frequency (blue) due to injection locking and synchronization to the external INJ signal (compare to measurements in Fig. 2a).

The term $i(\Omega_{rs}/2)(n_r - n_s)e^{i\delta_s t}$, in $\dot{\sigma}_{rs}$ acts as a coherent drive that modulates excited-state imbalance and feeds back into the nonlinear energy shift $E_{NL}$. This shifts the effective detunings $\Delta_{r,s}$ allowing the system's natural oscillation frequency and potential limit-cycle behaviors be tuned dynamically by $\Omega_{rs}$ and $\delta_s$.

Fig. 5 studies the numerical transient and spectral response of the $\Im(\sigma_{gr})$. Fig. 5a shows three cases for $\Im(\sigma_{gr})$: without OSC and in a stationary phase state (SS, red), in the OSC regime (blue), and OSC with added strong RF (green) to perturb the OSC. γ/2π=25.4 kHz is used consisting of spontaneous decay and transient time broadening. Perturbations when an RF is injected resulted in a gradual shift in the periodicity or frequency (green). Fig. 5b,c investigates the RF spectrum of $\Im(\sigma_{gr})$ under weak (b) and strong (c) injection (INJ) signals at detunings of +0.05 kHz, +0.1 kHz, and +0.75 kHz relative to the system's natural oscillation frequency (~23.45 kHz). In the weak-INJ regime, the drive has minimal effect on the oscillator dynamics but increases in response as it nears resonance. In contrast, strong INJ signals within a characteristic bandwidth (locking bandwidth) suppress the natural oscillation peak and possibly lead to synchronization with INJ frequency. Time-domain simulations (Fig. 6a) of injection locking reveal that once the INJ signal is activated, the natural oscillation rapidly diminishes and synchronizes to the drive, as shown by the spectrogram of $\Im(\sigma_{gr})$. The oscillation amplitude decreases with increasing $\Omega_{rs}$, and critical points (Fig. 6b) —defined by a 90% reduction in signal are used to mark onset of locking. These critical values shift with the detuning $\delta_s$, indicating that the locking range scales with the RF field strength. A linear relationship is observed between the locking bandwidth and $\Omega_{rs}$, consistent with $\Delta\omega = 2K\Omega_{rs}$ (K~0.014, Fig 6c).

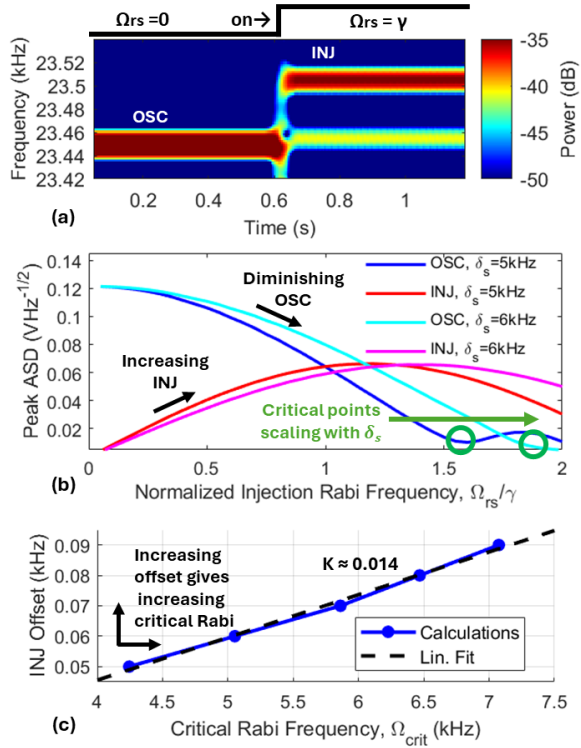

(a)
(b)
(c)

**Fig. 6**: Theoretical simulations of the INJ locking in time domain to study locking dynamics. (a) Spectrogram of the transients in $\Im[\sigma_{gr}]$ with a step function of the INJ signal (turned on at ~0.6s, Ωrs=γ). The OSC is observed to rapidly diminish in magnitude at the natural OSC frequency as the oscillatory dynamics rapidly synchronize and lock to the external INJ signal. (b) The amplitude spectral density (scales with magnitude) of the OSC is seen to diminish with increasing INJ magnitude (calculated as a function of Ωrs/γ). Critical points (green circles) identify locations where the the OSC magnitude reduces by ~90%. Calculation for δs=5kHz,6kHz observe a shift to the right (green arrow) for increasing offsets between the natural OSC and INJ frequencies, suggesting a scaling of locking range or bandwidth with field magnitude. (c) Calculation of critical points as a function of δs (based on 90% reduction in OSC magnitude) shows a linear relationship between INJ offset (half of locking range or bandwidth) and Rabi frequency (Δω=2KΩrs).

**Derivation of injection locking bandwidth**

Injection locking occurs when natural oscillatory dynamics synchronizes to an external drive through phase coupling[17]. The key dynamic is the evolution of the relative phase, whose equation governs locking behavior. In classical systems[17,18], the phase relationship obeys $\dot{\phi} = \Delta\omega - A\sin\phi$, leading to locking bandwidths proportional to the coupling strength. In the V-type atomic system considered, the excited-state coherence $\sigma_{rs}$ models the oscillatory dynamics, and its phase relative to the RF INJ determines the locking condition.

The key quantity is the phase difference $\phi(t)$ between the coherence and the driving RF field. The focus is on the dominant phase-coupling terms: the free evolution of $\sigma_{rs}$ and its direct RF drive at frequency $\delta_s$. The term $i\Omega(\sigma_{gs} - \sigma_{rg})/2$, which couples through intermediate coherences involving the ground state, is neglected here for simplicity, as it is subdominant in the locking regime:

$$\dot{\sigma}_{rs}^{(\text{eff})} \to -i(\Delta_r - \Delta_s - i\gamma)\sigma_{rs} + i\frac{\Omega_{rs}}{2}(n_r - n_s)e^{i\delta_s t}$$

To extract the phase dynamics, $\sigma_{rs}$ and its derivative $\dot{\sigma}_{rs}$ is expressed in a convenient form:

© 2025. California Institute of Technology. Government sponsorship acknowledged.

$$\sigma_{rs}^{(f)} = A(t)e^{i[\delta_s t + \phi(t)]}$$

$$\dot{\sigma}_{rs}^{(f)} = \left(\dot{A} + iA(\dot{\phi} + \delta_s)\right)e^{i(\delta_s t + \phi)}$$

where $A(t)$ is the amplitude and $\phi(t)$ is the phase difference with respect to the RF field. Setting $\dot{\sigma}_{rs}^{(\text{eff})} = \dot{\sigma}_{rs}^{(f)}$, substituting $\sigma_{rs} = \sigma_{rs}^{(f)}$, canceling the common exponential factor $e^{i\delta_s t}$, and isolating the imaginary parts (sensitive to phase) gives $[\Im(ie^{-i\phi}) = \cos\phi]$:

$$A(\dot{\phi} + \delta_s) = -(\Delta_r - \Delta_s)A + \frac{\Omega_{rs}}{2}(n_r - n_s)\cos\phi$$

Solving for $\dot{\phi}$, we obtain the effective phase equation:

$$\dot{\phi} = (\omega_{OSC} - \delta_s) + \frac{\Omega_{rs}}{2A}(n_r - n_s)\cos\phi$$

$$\dot{\phi} = \frac{\Delta\omega}{2} + K\Omega_{rs}\cos\phi$$

where $\omega_{OSC} \to \Delta_s - \Delta_r$ is the systems natural oscillation frequency, $\Delta\omega/2 = \omega_{OSC} - \delta_s$ is the offset between natural OSC and INJ frequency, and $K = (n_r - n_s)/2A$ is considered a forcing strength scaling factor. The final form mirrors Adler's equation in classical systems[17]. Injection locking occurs when $\Delta\omega < 2K\Omega_{rs}$. This condition ensures the phase equation admits a steady-state solution ($\dot{\phi} = 0$), indicating synchronization between the natural OSC and the RF drive. This defines the critical INJ Rabi frequency $\Omega_{rs}^{\text{crit}}$ as the critical value required for locking at a given detuning $\Delta\omega$, with locking bandwidth scaling linearly as $\Delta\omega_{\text{lock}} = 2K\Omega_{rs}^{\text{crit}}$.

**Experimental setup and approach**

The experiment utilizes a room-temperature Cesium vapor cell embedded with a parallel-plate capacitor, enabling application of low-frequency E-fields (RF INJ). Two interacting laser beams —a probe and a coupling laser—counter-propagate through the cell and overlap in the region between the capacitor plates. A B-field is used to induce competition between Rydberg sublevels at the upper states, breaking degeneracies and enabling Zeeman-induced dynamics. The system exhibits spontaneous oscillations in the population of excited Rydberg states, typically at ~11.2kHz, sustained by non-equilibrium feedback. A third probe beam is used as a reference to reduce technical noise.

The probe beam is tuned near the 6S$_{1/2}$ F=4 to 6P$_{3/2}$ F'=4 transition and locked to a hyperfine feature using Doppler-free saturation absorption spectroscopy. With a wavelength of approximately 852.36nm and a linewidth below 60kHz, the probe is detuned by $\Delta_p/2\pi$ =+22MHz from resonance and has a 1.1mm diameter (1/e²). It has a Rabi frequency of $\Omega_p/2\pi$ =13.97MHz. Polarization is controlled with a half-wave plate (λ/2) and manipulated using a calcite beam displacer to generate dual spatially separated probe beams (2.8mm offset) with orthogonal polarizations. These are fine-tuned with a quarter-wave plate (λ/4) to introduce and tune ellipticity to enhance observed OSC peak.

Balanced amplified photodetection (BAP) is used to monitor the probe signal, with the output digitized by a RF spectrum analyzer. To suppress higher-order spectral components and DC offsets, a combination of low-pass (20kHz low-pass) and DC-blocking filters is applied. This readout reveals (OSC) oscillations at ~11.2kHz in probe transmission driven by population dynamics among Rydberg sublevels.

The coupling laser is tuned to excite atoms from the intermediate state 6P$_{3/2}$ to the high-lying Rydberg level 76D$_{5/2}$ state. This laser operates near 508.81nm and is generated via second harmonic generation from a commercial external-cavity diode laser. Frequency stabilization is achieved using Pound-Drever-Hall locking to a medium-finesse cylindrical cavity housed in vacuum, narrowing the laser linewidth to below 100Hz (50mm diameter × 100mm length, finesse $F_c$~15k). The coupling beam has a 1.3mm diameter (1/e²), operates with a detuning $\Delta_c/2\pi$ =-30MHz and a Rabi frequency of approximately $\Omega_c/2\pi$ =0.85MHz. The coupler beam polarization is tuned with a quarter-wave plate (λ/4) to introduce ellipticity, key to driving multiple sublevels and to enhance observed OSC peak.

A weak magnetic field (B=4G) is applied along the laser axis to lift the degeneracy of closely spaced Rydberg Zeeman sublevels, particularly between the J=3/2 and J=5/2 manifolds, where the latter exhibits stronger dipole coupling. At B=4G in our system, the induced Zeeman shifts lead to inhomogeneous broadening and competing transitions, driving dynamical instabilities and resulting in sustained oscillations (OSC at ~11.2kHz) in probe transmission. The coupling laser's elliptically polarized light—adjusted via quarter-wave plates—is essential for enabling simultaneous population of multiple Zeeman sublevels, which supports persistent oscillatory dynamics arising from collective many-body interactions[6].

An RF INJ E-field (near OSC frequencies), aligned with the capacitor plates, modulates the energy separation between sublevels participating in the OSC dynamics. This field serves as an external INJ RF drive, allowing investigation of frequency pulling and injection locking.

For fixed RF frequencies (e.g., $f_{RF}$=10.25kHz), increasing the electric field amplitude pulls the observed OSC frequency toward the drive frequency. As the field strength exceeds a threshold, the system abruptly synchronizes with the RF source — exhibiting classical injection locking behavior. Conversely, when the drive frequency is swept at constant field magnitude (e.g., E=4mV/cm), the system's oscillation frequency is gradually pulled toward resonance until locking occurs, revealing a characteristic injection locking bandwidth (see Fig. 1e,f).

Additional details of experimental systems and setup are presented in Supplementary Section 2.

### Data availability

The following are available as source data: 1) Colormap (2D array) dataset of INJ field magnitude vs frequency showing dependence of IL as a function critical INJ field magnitude ($f_{INJ}$= 10.25kHz, IL occurs at ~5mV/cm) in Fig. 2d; 2) Colormap (2D array) with E-field held constant at 4 mV/cm, showing a gradual pulling (IP) as the RF field frequency is slowly tuned closer to the natural OSC frequency in Fig. 3a. Higher order harmonics of the OSC is also IP towards the INJ frequency; 3) Colormap (2D array) of INJ frequency sweep from about 12.3-10kHz (4.2mV/cm) in Fig. 4a. IP is observed to pull OSC towards INJ and locking occurs at about ~11.8kHz. The system remains IL (INJ locked) between ~11.8-10.6kHz. All other data are available upon reasonable request.

### References

1. Diehl, S., Micheli, A., Kantian, A. et al. Quantum states and phases in driven open quantum systems with cold atoms. Nature Phys 4, 878–883 (2008). https://doi.org/10.1038/nphys1073
2. Marcuzzi, M., Levi, E., Diehl, S., Garrahan, J. P. & Lesanovsky, I. Universal nonequilibrium properties of dissipative Rydberg gases. Nat. Phys. 113, 210401 (2014). DOI: 10.1103/PhysRevLett.113.210401.
3. Lee TE, Häffner H, Cross MC. Antiferromagnetic phase transition in a nonequilibrium lattice of Rydberg atoms. *Phys Rev A.* 2011;84(3):031402. doi:10.1103/PhysRevA.84.031402.
4. Wadenpfuhl K, Adams CS. Emergence of synchronization in a driven-dissipative hot Rydberg vapor. Phys Rev Lett. 2023;131(14):143002. doi:10.1103/PhysRevLett.131.143002.
5. Dongsheng Ding et al., Ergodicity breaking from Rydberg clusters in a driven-dissipative many-body system. Sci. Adv. 10, eadl5893 (2024). DOI:10.1126/sciadv.adl
6. Wu, X., Wang, Z., Yang, F. et al. Dissipative time crystal in a strongly interacting Rydberg gas. Nat. Phys. 20, 1389–1394 (2024). https://doi.org/10.1038/s41567-024-02542-9
7. Wilczek, F. Quantum time crystals. Phys. Rev. Lett. 109, 160401 (2012).
8. Krishna, M., Solanki, P., Hajdušek, M. & Vinjanampathy, S. Measurement-Induced Continuous Time Crystals. Phys. Rev. Lett. 130, 150401 (2023).
9. Carollo, F. & Lesanovsky, I. Exact solution of a boundary time-crystal phase transition: time-translation symmetry breaking and non-Markovian dynamics of correlations. Phys. Rev. A 105, L040202 (2022).
10. Iemini, F., Russomanno, A., Keeling, J., Schirò, M., Dalmonte, M. & Fazio, R. Boundary time crystals. Phys. Rev. Lett. 121, 035301 (2018).
11. Nozières, P. Time crystals: Can diamagnetic currents drive a charge density wave into rotation? Europhys. Lett. 103, 57008 (2013).
12. Bakker, L. R., Bahovadinov, M. S., Kurlov, D. V., Gritsev, V., Fedorov, A. K. & Krimer, D. O. Driven-dissipative time crystalline phases in a two-mode bosonic system with Kerr nonlinearity. Phys. Rev. Lett. 129, 250401 (2022).
13. Nie, X. & Zheng, W. Mode softening in time-crystalline transitions of open quantum systems. Phys. Rev. A 107, 033311 (2023).
14. Keßler, H., Cosme, J. G., Hemmerling, M., Mathey, L. & Hemmerich, A. Emergent limit cycles and time crystal dynamics in an atom-cavity system. Phys. Rev. A 99, 053605 (2019).
15. Buča, B., Tindall, J. & Jaksch, D. Non-stationary coherent quantum many-body dynamics through dissipation. Nat. Commun. 10, 1730 (2019)
16. Dogra, N., Landini, M., Kroeger, K., Hruby, L., Donner, T. & Esslinger, T. Dissipation-induced structural instability and chiral dynamics in a quantum gas. Science 366, 1496–1499 (2019).
17. R. Adler, "A Study of Locking Phenomena in Oscillators," Proceedings of the IRE, vol. 34, no. 6, pp. 351–357, 1946.
18. Kuramoto, Y. (1975). Self-entrainment of a population of coupled nonlinear oscillators. In H. Araki (Ed.), International Symposium on Mathematical Problems in Theoretical Physics, vol. 39, pp. 420–422. Springer.
19. Roulet, A. & Bruder, C. Quantum synchronization and entanglement generation. Phys. Rev. Lett. 121, 063601 (2018).
20. Zhang, L., Wang, Z., Wang, Y., Zhang, J., Wu, Z., Jie, J. & Lu, Y. Quantum synchronization of a single trapped-ion qubit. Phys. Rev. Res. 5, 033209 (2023).
21. L. J. Paciorek, "Injection locking of oscillators," Proc. IEEE, vol. 53, no. 11, pp. 1723-1727, 1965.
22. Pikovsky, A., Rosenblum, M. & Kurths, J. Synchronization: A Universal Concept in Nonlinear Sciences. (Cambridge Univ. Press, 2001).
23. Strogatz, S. H. From Kuramoto to Crawford: exploring the onset of synchronization in populations of coupled oscillators. Physica D 143, 1–20 (2000).
24. Acebrón, J. A. et al. The Kuramoto model: A simple paradigm for synchronization phenomena. Rev. Mod. Phys. 77, 137–185 (2005).

**Figure Legends:**

**Fig. 1: Experimental protocol, energy diagram, mean-field treatment simulations, and experimental observations. (a) Counter-propagating probe and coupler laser beams are directed through a room-temperature Cesium vapor cell. Inside the cell, a parallel-plate capacitor enables the application of low-frequency electric fields (E-field). A static magnetic field (B-field) is applied to induce mode competition among the excited Rydberg state sublevels, resulting in limit cycle oscillations (OSC). To reduce classical noise, a reference probe beam is employed. (b) The experimental setup follows a ladder-type configuration, where a probe laser at $\lambda_p$~852nm drives the ground state $|g\rangle=6S_{1/2}$ to the intermediate excited state $|e\rangle=6P_{3/2}$, detuned by $\Delta_p/2\pi$=22MHz. A coupler laser at $\lambda_c$~508nm then excites atoms from $|e\rangle$ to a high-lying Rydberg state $76D_{5/2}$, detuned by $\Delta_c/2\pi$=30MHz. A B=4G field is applied to induce Zeeman shifts**

and drive competition between the Rydberg sublevels. A low-frequency E-field modulates the energy difference between two nearby Rydberg states |r⟩ and |s⟩, with the detuning δs defined relative to their energy separation. For theoretical modeling, a V-type three-level system (gray) is used to approximate the dynamics. (c) Time-dependent numerical simulations of the V-type model shows that the OSC frequency is dynamically frequency locked to injection (INJ) signal when turned on. (d) Numerical simulation with varying INJ Rabi frequency shows a peak INJ lock close to $\Omega rs/\gamma \sim 1$, where $\gamma$ is the linewidth of the upper states. (e) Experimental observations for fixed RF E-field frequency ($f_{INJ}$=10.25kHz) shows that as E-field magnitude is increased, the OSC (naturally at ~11.2kHz) is gradually pulled towards the INJ signal until at abruptly locks to the INJ frequency. This is the characteristic injection pulling (IP) in oscillatory dynamics. (f) For a fixed E-field magnitude (E=4mV/cm), a similar gradual IP pulling is observed when the RF E-field frequency is gradually moved towards the natural OSC frequency.

Fig. 2: (a-c) Spectral snapshots of OSC dynamics under strong and weak RF injection (INJ). The INJ frequency ($f_{INJ}$) is gradually moved towards the natural OSC frequency. (a) E=4.2mV/cm. OSC at about 11.2kHz is seen to gradually shift towards INJ, locking at ~11.8kHz. The locking is referred to as INJ lock (IL). Within an effective bandwidth from the OSC, referred to as the locking bandwidth, the signal remains locked. (b) E=0.83mV/cm. OSC at about 11.2kHz remains at the natural OSC frequency and does not shift towards INJ, demonstrating that weak INJ do not result in IP (pulling). (c) E=5.8mV/cm. OSC at about 11.2kHz rapidly shifts towards INJ, locking at ~12.3kHz. The observations confirm that IP is dependent on the strength of the INJ signal, and that the locking bandwidth is dependent on INJ field magnitude. (d-e) Colormap datasets of INJ field magnitude vs spectral frequency showing dependence of IL as a function critical INJ field magnitude. Natural OSC frequencies at about 11.2kHz in all cases. (d) $f_{INJ}$= 10.25kHz. IL occurs at ~5mV/cm. (e) $f_{INJ}$= 10.5kHz. IL occurs at ~4mV/cm. (f) $f_{INJ}$= 10.75kHz. IL occurs at ~2.5mV/cm. The critical field magnitude to IL the OSC increases as a function of separation between INJ and OSC frequencies.

Fig. 3: (a) With E-field held constant at 4 mV/cm, a gradual pulling (IP) occurs as the RF field frequency is slowly tuned closer to the natural OSC frequency. Higher order harmonics of the OSC is also IP towards the INJ frequency. (b) In a step-on regime, where the INJ (4mV/cm) is turned on abruptly and is within the locking bandwidth, the OSC rapidly locks to the INJ. (c) Natural OSC dynamics show frequency drifts and have larger 3dB (FWHM) bandwidth (error bars) compared to after injection locking, where the dynamics and FWHM follow the characteristics of the INJ.

Fig. 4: Locking, unlocking, and bandwidths. (a) INJ frequency is swept from about 12.3-10kHz (4.2mV/cm). IP is observed to pull OSC towards INJ and locking occurs at about ~11.8kHz. The system remains IL (INJ locked) between ~11.8-10.6kHz (locking bandwidth is visually estimated here at about ~1.2kHz). (b) Locking bandwidth as a function of critical electric field, extracted by performing linear fits to the readout OSC frequency shift versus applied field for each injection frequency. The critical field is defined as the intercept point of the fit, corresponding to where the OSC is estimated to linearly intercept the INJ frequency. Experimental measurements (blue) are compared with theoretical predictions for two values of the scaling factor α, where $\Delta\omega=\alpha E$, α (kHz/mVcm$^{-1}$) and $\Delta\omega$ is the locking bandwidth (see Methods). The estimated $\Delta\omega$ is reduced relative to visual estimation due to linear fit estimation of intercept points.

Fig. 5: Time-dependent numerical simulations of the V-type atom with a mean field treatment. (a) Transient response of the $\Im[\sigma_{gr}]$ that relates to probe transmission showing rapid decay (red) outside the OSC regime (stationary phase, SS), sustained oscillations (blue) in the OSC region, and perturbations when an RF is injected resulting in a gradual shift in the periodicity or frequency (green). (b-c) The spectrum of $\Im[\sigma_{gr}]$ is studied for weak and strong INJ signals with frequencies $f_{INJ}$ at +0.05kHz, +0.75kHz, and +0.1kHz relative to natural OSC frequencies. A reference line (purple) is included to identify the case where $\Omega rs=0$ (no INJ signal). (b) Weak INJ signals do not perturb the OSC dynamics (at about ~23.45kHz) and increases in magnitude as it approaches the OSC (compare to measurements in Fig. 2b). (c) Strong INJ signals perturb the OSC. When INJ signals are within the locking bandwidth and above a critical magnitude, they result in a rapidly diminishing magnitude of the OSC signal at the natural OSC frequency (blue) due to injection locking and synchronization to the external INJ signal (compare to measurements in Fig. 2a).

Fig. 6: Theoretical simulations of the INJ locking in time domain to study locking dynamics. (a) Spectrogram of the transients in $\Im[\sigma_{gr}]$ with a step function of the INJ signal (turned on at ~0.6s, $\Omega rs=\gamma$). The OSC is observed to rapidly diminish in magnitude at the natural OSC frequency as the oscillatory dynamics rapidly synchronize and lock to the external INJ signal. (b) The amplitude spectral density (scales with magnitude) of the OSC is seen to diminish with increasing INJ magnitude (calculated as a function of $\Omega rs/\gamma$). Critical points (green circles) identify locations where the the OSC magnitude reduces by ~90%. Calculation for $\delta s$=5kHz,6kHz observe a shift to the right (green arrow) for increasing offsets between the natural OSC and INJ frequencies, suggesting a scaling of locking range or bandwidth with field magnitude. (c) Calculation of critical points as a function of $\delta s$ (based on 90% reduction in OSC magnitude) shows a linear relationship between INJ offset (half of locking range or bandwidth) and Rabi frequency ($\Delta\omega=2K\Omega rs$).

### Acknowledgements

The author would like to acknowledge discussions with P. Mao and D. Willey at JPL (Jet Propulsion Laboratory, California Institute of Technology), A. Artusio-Glimpse, N. Prajapati, C. Holloway, and M. Simons at NIST (National Institute of Standards and Technology), and K. Cox, D. Meyer, and P. Kunz at ARL (Army Research Laboratory) as part of the NASA Instrument Incubator Program on Rydberg Radars. The research was carried out at the Jet Propulsion Laboratory, California Institute of Technology, under a contract with the National Aeronautics and Space Administration (80NM0018D0004), through the Instrument Incubator Program's (IIP) Instrument Concept Development (Task Order 80NM0022F0020).

### Author contributions

D.A conceived of the experiment and study reported, configured the atomic systems to include lasers and locking systems, and collected and processed all data reported in the text and figures. D.A. also developed all modeling, theoretical derivations and numerical simulations used or reported.

### Additional information

The author declares no competing interest.

**Supplementary information**

**Supplementary Section 1**

**Features of a Rydberg dissipative time crystal (DTC)**

Dissipative time crystals (DTCs) are driven many-body systems that exhibit persistent, intrinsic oscillations arising from nonlinear interactions, rather than from direct harmonic driving. A Rydberg DTC is characterized by four defining features: (1) Spontaneous time-translation symmetry breaking (TTSB), where oscillations emerge intrinsically rather than being imposed externally[25]; (2) Long-range temporal coherence, marked by stable, phase-locked oscillations persisting over time[26]; (3) Robustness to perturbations, with oscillatory dynamics surviving in the presence of external noise, detunings, or signal injection[27]; (4) Many-body interactions, which drive collective behavior beyond simple two-body dynamics[28].

In our room-temperature system, these features emerge when atoms are excited to high principal quantum numbers (e.g., n=76). At this regime, the energy spacing between Rydberg $nD_{3/2}$ and $nD_{5/2}$ manifolds becomes small, and the stronger coupling of the J=5/2 state enhances interaction-induced nonlinearities[6]. This facilitates a non-equilibrium dynamical phase transition (see Extended Data Fig. 4 in Ref.[6]), particularly under weak magnetic B-fields. As the B-field increases, driven Zeeman splitting induces inhomogeneous broadening, yet within a specific range (4-5G in our system), closely spaced sublevels enable coherent competition among multiple Rydberg states, driving sustained oscillatory behavior that can be observed in probe transmission.

The polarization of the coupling laser further enables redistribution across Zeeman sublevels. In particular, elliptically polarized light, tuned via quarter-wave plates, maximizes population mixing and supports mode competition essential to the oscillatory phase[6].

The observed dynamics exhibit signatures of a genuine many-body effect: oscillations persist far beyond what would be expected from isolated atom pairs or collisional broadening. Instead, interactions among many Rydberg sublevels stabilize collective oscillations through nonlinear feedback mechanisms shaped by the magnetic field and competition. These properties confirm all four DTC criteria in our system. Spontaneous limit-cycle oscillations demonstrate TTSB[25-27]; temporal coherence is evidenced by non-decaying autocorrelations[6]; robustness is observed through resilience to noise[6]; and interaction-driven dynamics are supported by nonlinear detuning shifts and interlevel competition[5,6]. Together, these observations establish the formation of a Rydberg DTC in a room-temperature, driven-dissipative ensemble.

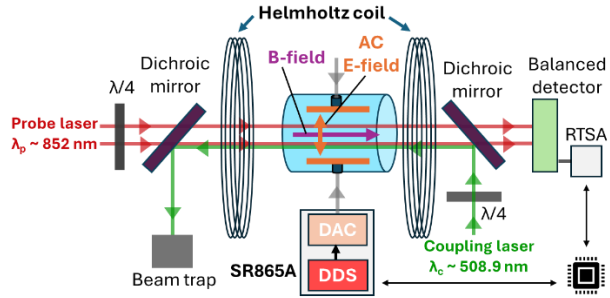

**Extended Data Fig. 1:** The experimental configuration includes a pair of Helmholtz coils, each with a diameter of 30cm and spaced 15cm apart, providing a uniform B-field of 4G. A cylindrical vapor cell, measuring 56.6mm in length and 35mm in diameter, is positioned at the center of the coil system. Inside the cell, a pair of embedded plates separated by 15mm enables the generation of a controllable RF electric field. Two laser systems are used to drive the atomic transitions. The probe laser, operating near 852nm, excites the transition from the $6S_{1/2}$ F=4 ground state to the $6P_{3/2}$ F'=4 excited state. This diode laser (Toptica DL Pro) is frequency-stabilized via saturated absorption spectroscopy. The coupling laser, a frequency-doubled system (Toptica TA-MSHG PRO), addresses the upper transition from $6P_{3/2}$ F'=4 to the highly excited $76D_{5/2}$ Rydberg state. It is frequency-locked to a medium-finesse optical cavity with a finesse of approximately 15k. A Stanford Research Systems SR865A lock-in amplifier is employed to generate RF signals programmatically, which are used to drive the electric field plates. This lock-in is used solely for signal generation and not for signal detection.

**Supplementary Section 2**

**Details of experimental setup**

Extended Data Fig. 1 outlines the core layout of the experiment. Additional system-level specifications are detailed below.

A uniform magnetic field of 4 G is generated using a matched pair of Helmholtz coils (3B Scientific, 300 mm diameter, 320 turns per coil) positioned 15 cm apart. The vapor cell, made of Pyrex and measuring 56.6 mm in length and 35 mm in diameter, contains two internal stainless-steel electrodes (18 × 45 mm, 15 mm separation) that form a parallel-plate geometry for applying tunable electric fields. The cell is uncoated and filled with pure Cesium vapor at room temperature, with no buffer gas.

RF signals are delivered to the electrodes using a Stanford Research Systems SR865A digital lock-in amplifier. This instrument is chosen for its low-noise output and high-resolution amplitude control and is used exclusively for signal generation.

The probe laser (Toptica DL Pro) at 852nm, is stabilized using saturated absorption spectroscopy and used to interrogate the ground-state transition. Transmission is measured using a Thorlabs PDB250A2 balanced photodetector. The signal path includes a DC block and an in-line 20 kHz low-pass filter to suppress unwanted components. A Keysight MXA N9021 spectrum analyzer is used for data acquisition in swept modes. Spectral data are collected with a resolution bandwidth of 10 Hz.

The coupling laser (Toptica TA-MSHG PRO) is generated via second harmonic generation of a ~1016 nm fundamental beam. The system employs a tapered amplifier followed by a resonantly enhanced nonlinear crystal to generate ~508.8 nm light. Frequency stabilization is achieved through Pound-Drever-Hall locking to a custom cylindrical optical cavity (Stable Laser Systems; 50 mm diameter, 100 mm length, finesse ~15k) housed in vacuum. A HF-ANGSTROM WS/7 wavemeter monitors both probe and coupling laser wavelengths for referencing and diagnostics.

### Supplementary Section 3
**Population evolution and coherence equations**

The operator $\hat{O}$ is evolves as $\partial_t \langle \hat{O} \rangle = i \langle [\hat{H}, \hat{O}] \rangle + \mathcal{L}^a[\hat{O}]$, where the Lindblad term accounts for the decay of the Rydberg states: $\mathcal{L}^a[\hat{O}] = \sum_i \gamma (2 \hat{\sigma}_{\alpha g}^i \hat{O} \hat{\sigma}_{g\alpha}^i - \{\hat{n}_\alpha^i, \hat{O}\})$ ($\alpha = r, s$). Taking into account the additional coupling between the excited states |r⟩ and |s⟩, the first-moment equations of motion reflect the dynamics of population exchange governed by excitation, spontaneous decay, and the interaction between the excited states:

$$\frac{d}{dt}\langle \hat{n}_r^i \rangle = i\frac{\Omega}{2}(\langle \hat{\sigma}_{gr}^i \rangle - \langle \hat{\sigma}_{rg}^i \rangle) - \gamma \langle \hat{n}_r^i \rangle$$
$$+ i\frac{\Omega_{rs}}{2}(\langle \hat{\sigma}_{rs}^i e^{i\delta_s t}\rangle - \langle \hat{\sigma}_{sr}^i e^{-i\delta_s t}\rangle)$$

$$\frac{d}{dt}\langle \hat{n}_s^i \rangle = i\frac{\Omega}{2}(\langle \hat{\sigma}_{gs}^i \rangle - \langle \hat{\sigma}_{sg}^i \rangle) - \gamma \langle \hat{n}_s^i \rangle$$
$$- i\frac{\Omega_{rs}}{2}(\langle \hat{\sigma}_{rs}^i e^{i\delta_s t}\rangle - \langle \hat{\sigma}_{sr}^i e^{-i\delta_s t}\rangle),$$

and by the equations governing coherence dynamics, which are shaped by interactions and Rabi oscillations, now including extra terms arising from the influence of $\Omega_{rs}$ induced by the RF injected field:

$$\frac{d}{dt}\langle \hat{\sigma}_{gr}^i \rangle = i\frac{\Omega}{2}(2\langle \hat{n}_r^i \rangle + \langle \hat{n}_s^i \rangle + \langle \hat{\sigma}_{sr}^i \rangle - 1)$$
$$+ i\left(\Delta_r - \sum_{j\neq i} V_{ij}(\langle \hat{n}_r^j \rangle + \langle \hat{n}_s^j \rangle) + i\frac{\gamma}{2}\right)\langle \hat{\sigma}_{gr}^i \rangle$$

$$\frac{d}{dt}\langle \hat{\sigma}_{gs}^i \rangle = i\frac{\Omega}{2}(2\langle \hat{n}_s^i \rangle + \langle \hat{n}_r^i \rangle + \langle \hat{\sigma}_{rs}^i \rangle - 1)$$
$$+ i\left(\Delta_s - \sum_{j\neq i} V_{ij}(\langle \hat{n}_r^j \rangle + \langle \hat{n}_s^j \rangle) + i\frac{\gamma}{2}\right)\langle \hat{\sigma}_{gs}^i \rangle$$

$$\frac{d}{dt}\langle \hat{\sigma}_{rs}^i \rangle = i\frac{\Omega}{2}(\langle \hat{\sigma}_{gs}^i \rangle - \langle \hat{\sigma}_{rg}^i \rangle) - i(\Delta_r - \Delta_s - i\gamma)\langle \hat{\sigma}_{rs}^i \rangle$$
$$+ i\frac{\Omega_{rs}}{2}(\langle \hat{n}_r^i \rangle - \langle \hat{n}_s^i \rangle)e^{i\delta_s t}.$$

### Supplementary Section 4
**Repeatability of injection locking in Rydberg DTC**

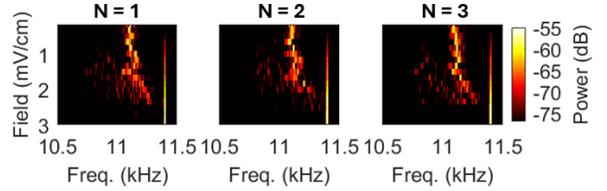

**Extended Data Fig. 2:** Colormap datasets of INJ field magnitude vs spectral frequency showing IP (pulling) followed by IL (abrupt locking). Natural OSC frequencies is about 11.1kHz in all cases, while INJ frequency is $f_{INJ}$= 11.4kHz. Data is collected over three separate days (N = 1,2,3) to show repeatability of IP dynamics.

To assess the stability and reproducibility of injection pulling (IP) and locking (IL) dynamics, repeated measurements were performed on three separate days under nominally identical conditions. Extended Data Fig. 2 presents colormap datasets showing the spectral response as a function of applied RF electric field for each dataset (N = 1, 2, 3). In all cases, the natural oscillation (~11.1kHz) is gradually pulled toward the injected signal at 11.4kHz, followed by abrupt synchronization. Injection locking consistently occurs at a field of 2.5 ± 0.1mV/cm, identified by the point where the natural oscillation converges with the injected frequency. This consistency demonstrates the high repeatability of the locking threshold across independent experimental runs.

### Supplementary Section 5
**Field-Frequency maps as a function of INJ frequency**

To fully characterize the injection pulling (IP) and locking (IL) behavior, it is important to study how the system responds to a range of injected RF frequencies. Mapping the spectral dynamics as a function of both electric field strength and injection frequency allows visualization of IP behavior when the injected frequency lies above or below the natural oscillation frequency (OSC). This provides insight into the symmetry and extent of entrainment across driving configurations. Extended Data Fig. 3 presents a series of colormaps showing probe transmission spectra for different fixed INJ RF frequencies, each as a function of applied E-field magnitude. The injected frequency in each case is indicated by a vertical bright trace (white arrow). As the E-field increases, the natural oscillation signal is pulled toward the injected frequency in all cases, regardless of whether the injection frequency is higher or lower than the OSC. For injection frequencies closer to the natural OSC, IL occurs at lower field thresholds. Additionally, higher-order harmonics of the OSC are also pulled toward the drive frequency, indicating that the nonlinear structure of the oscillation is itself entrained by the injected field. This suggests that injection locking not only stabilizes the fundamental mode but suppresses higher-order dynamics, effectively cleaning the spectrum and enhancing coherence.

### Additional References for Supplementary Sections

25. Else, D. V., Bauer, B. & Nayak, C. Floquet time crystals. Phys. Rev. Lett. 117, 090402 (2016). https://doi.org/10.1103/PhysRevLett.117.090402
26. Sacha, K. & Zakrzewski, J. Time crystals: a review. Rep. Prog. Phys. 81, 016401 (2017). https://doi.org/10.1088/1361-6633/aa8b38
27. Yao, N. Y., Potter, A. C., Potirniche, I.-D. & Vishwanath, A. Discrete time crystals: rigidity, criticality, and realizations. Phys. Rev. Lett. 118, 030401 (2017). https://doi.org/10.1103/PhysRevLett.118.030401
28. Khemani, V., Lazarides, A., Moessner, R. & Sondhi, S. L. Phase structure of driven quantum systems. Phys. Rev. Lett. 116, 250401 (2016). https://doi.org/10.1103/PhysRevLett.116.250401

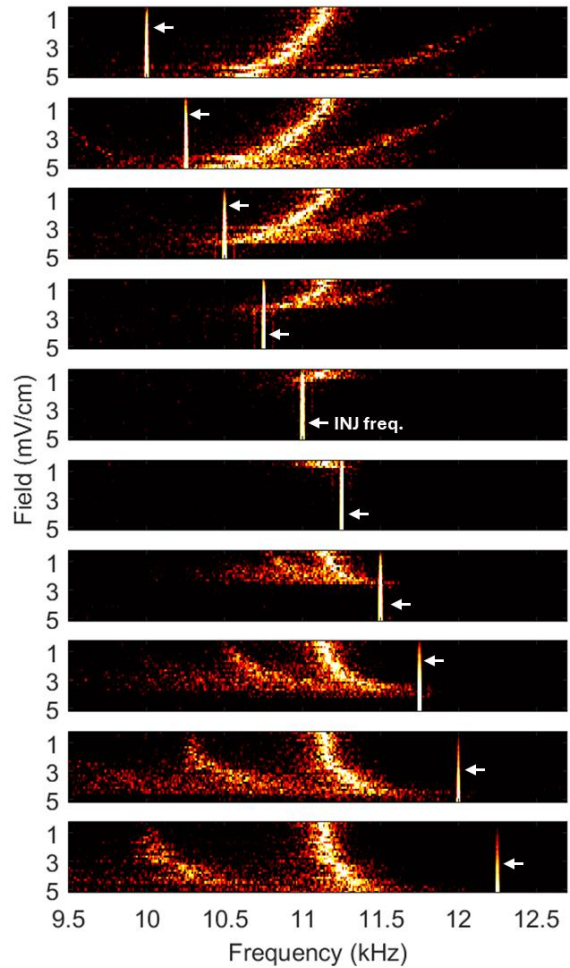

**Extended Data Fig. 3:** Colormap datasets collected as part of a data collection campaign, with E-field magnitude <5mV/cm and for spectral frequencies between 9.5-12.5kHz. Each tile corresponds to a different INJ frequency (denoted by vertical varying magnitude signal signature and a white arrow). In each case, the OSC dynamics and frequency, along with its higher order harmonics are IP (injection pulled) towards the INJ frequency, regardless of whether the INJ frequency is lower or higher than the OSC frequency. INJ frequencies closer to natural OSC frequencies result in IL (injection locking) at lower E-field magnitudes.